\documentclass[useAMS,usenatbib]{mn2e}
\usepackage[british]{babel}	
\usepackage{amsmath}
\usepackage[T1]{fontenc}
\usepackage{verbatim}
\usepackage{graphicx,subfigure,float}
\usepackage[font=small,hang]{caption}
\usepackage{txfonts}
\usepackage{upgreek}
\usepackage[breaklinks]{hyperref}
\usepackage{tikz}
\usetikzlibrary{calc,trees,positioning,arrows,chains,shapes.geometric,%
    decorations.pathreplacing,decorations.pathmorphing,shapes,%
    matrix,shapes.symbols}

\def\hi{\ifmmode{\rm HI}\else{H\/{\sc i}}\fi} 
\def\ovii{\ifmmode{\rm OVII}\else{O\/{\sc vii}}\fi}
\def\oviii{\ifmmode{\rm OVIII}\else{O\/{\sc viii}}\fi}
\newcommand {\kms} {\,{\rm km\,s}^{-1}}

\newcommand {\mo}{{\rm M}_\odot}

\newcommand {\moyr}{\,{\rm M_\odot\,\rm yr}^{-1}}

\title[Efficiency of gas cooling and accretion at the disc-corona interface]{Efficiency of gas cooling and accretion at the disc-corona interface}
\author[L. Armillotta, F. Fraternali and F. Marinacci]{L. Armillotta$^{1}$\thanks{E-mail: lucia.armillotta@unibo.it}, F. Fraternali$^{1,2}$, F. Marinacci$^{3}$\\
$^{1}$Department of Physics and Astronomy, University of Bologna, 6/2, Viale Berti Pichat, 40127 Bologna, Italy\\
$^{2}$Kapteyn Astronomical Institute, Postbus 800, 9700 AV Groningen, The Netherlands\\
$^{3}$Department of Physics, Kavli Institute for Astrophysics and Space Research, Massachusetts Institute of Technology, Cambridge, MA 02139, USA}

\begin{document}


\pagerange{\pageref{firstpage}--\pageref{lastpage}} \pubyear{2016}

\maketitle

\label{firstpage}

\begin{abstract}
In star-forming galaxies, stellar feedback can have a dual effect on the circumgalactic medium both suppressing and stimulating gas accretion. The trigger of gas accretion can be caused by disc material ejected into the halo in the form of fountain clouds and by its interaction with the surrounding hot corona. Indeed, at the disc-corona interface, the mixing between the cold/metal-rich disc gas ($T \lesssim 10^4$ K) and the hot coronal gas ($T \gtrsim 10^6$ K) can dramatically reduce the cooling time of a portion of the corona and produce its condensation and accretion.
We studied the interaction between fountain clouds and corona in different galactic environments through parsec-scale hydrodynamical simulations, including the presence of thermal conduction, a key mechanism that influences gas condensation. Our simulations showed that the coronal gas condensation strongly depends on the galactic environment, in particular it is less efficient for increasing virial temperature/mass of the haloes where galaxies reside and it is fully ineffective for objects with virial masses larger than $10^{13}\mo$. This result implies that the coronal gas cools down quickly in haloes with low-intermediate virial mass ($M_\mathrm{vir} \lesssim 3\times10^{12}\mo$) but the ability to cool the corona decreases going from late-type to early-type disc galaxies, potentially leading to the switching off of accretion and the quenching of star formation in massive systems.
\end{abstract}


\section{Introduction}
\label{Introduction}

Current cosmological models predict that star-forming galaxies like the Milky Way are embedded in hot gas atmospheres at virial temperature ($T \gtrsim 10^6$ K), the so-called `cosmological coronae', extending out to hundreds of kpc from the galaxy center \citep[e.g.][]{Fukugita&Peebles06}. On the basis of a combination of big bang nucleosynthesis theory and observations of the cosmic microwave background, it is widely accepted that these coronae contain a significant fraction of the warm-hot component of the missing baryons (WHIM) in the Universe \citep[e.g.][]{Fukugita&Peebles06, Komatsu+09, Shull+12}. Unfortunately the X-ray surface brightness of these coronae is considered too faint to be detected with the current generation of instruments \citep{Bregman07}. To date, haloes of hot gas have been observed around some massive spiral galaxies. In the giant spirals NGC 1961, UGC 12591 and NGC 266, X-ray emission has been detected at more than 50 kpc from the center, indicating the presence of extended structures of hot gas \citep{Dai+12, Bogdan+13, Anderson+16}. The mass of these coronae is comparable with the disc baryonic mass ($\lesssim 10^{11} \mo$), accounting for $10 - 50 \%$ of the missing baryons associated to those galaxies. 

In the Milky Way, the existence of a hot corona was originally hypothesized by \citet{Spitzer56} as a medium to provide pressure confinement to the High-Velocity Clouds \citep[HVCs,][]{Wakker&vanWoerden97}. Most of the evidence that we have is indirect and comes from absorption UV spectroscopy \citep{Sembach+03}, head-tail structure of several HVCs \citep{Putman+11}, rotation measure of pulsars in the Magellanic Stream \citep{Anderson&Bregman10} and ram pressure stripping of dwarf galaxies in the Local Group \citep{Grcevich&Putman09, Gatto+13}. Recently, the presence of a hot medium around the Galactic disc was detected through \ovii\ and \oviii\ emission lines in the Milky way's soft X-ray background \citep{Miller&Bregman15}.

Cosmological coronae seem to be a significant reservoir of gas that could be accreted by star-forming galaxies to sustain star formation at the current observed rates. Indeed a long standing problem in the evolution of star-forming galaxies like the Milky Way is how they keep accreting gas from the environment to feed their star formation. The star formation rate of these galaxies has mildly declined throughout their life time ($\sim 10$ Gigayears) \citep[e.g.][]{Aumer&Binney09, Fraternali&Tomassetti12}. In addition, it appears that the gas content of these galaxies has remained approximately unchanged throughout the Hubble time \citep[e.g.][]{Bauermeister+10, Zafar+13}. Typically, the mass of gas contained in the thin disc can sustain the process of star formation for a few gigayears only and thus, at any given cosmic epoch, spiral galaxies need some supply of external gas to be brought into the disc, at a rate of $\sim 1 \moyr$, that compensates the conversion of gas into stars \citep[e.g][]{Sancisi+08}. This gas is thought to be metal poor in order to fulfil chemical evolution models of spiral galaxies \citep[e.g.][]{Pagel09,Matteucci12} and to explain the observed metallicity gradients \citep[e.g.][]{Cavichia+14,Pezzulli&Fraternali16}.

For decades, HVCs have been considered as the main candidates for accreting cold and low-metallicity material onto the Milky Way, but the most recent estimates of their accretion indicate that the accrection rate is merely $\sim 0.08 \moyr$ \citep{Putman+12}, more than an order of magnitude lower than the Galactic star formation rate. The situation does not change in nearby galaxies where the amount of extragalactic cold gas accretion seems to account only for $\sim 10\%$ of the star formation rate \citep[e.g.][]{Sancisi+08, DiTeodoro&Fraternali14}. Hence, accretion from cosmological coronae seems to be the only viable possibility. However this coronal gas is very hot and rarefied and it needs to cool in order to become available for star formation \citep[e.g.][]{Miller&Bregman15}. 

How gas cooling and accretion takes place is a matter of debate. The hypothesis originally put forward to explain how the coronal gas might cool and collapse on the disc assumes the creation of thermal instabilities in the corona that lead to the development of cold clouds \citep[e.g.][]{Maller&Bullock04, Kaufmann+06}. However, later studies demonstrated that in galaxies similar to the Milky Way thermal instability in coronae is damped by the combined effect of heat conduction and buoyancy: cold clouds smaller than 10 kpc can form only farther than 100 kpc from the disc \citep{Binney+09, Nipoti&Posti13}, in contradiction with the distances determined for HVCs \citep[$\sim 5-20$ kpc,][]{Wakker01, Wakker07}. Finally, both adaptive mesh refinement cosmological simulations of Milky Way-like galaxies and smoothed particle hydrodynamical simulations with an appropriate treatment of phase mixing show that spontaneous cooling of the corona does not occur \citep{Joung+12, Hobbs+13}. \citet{Joung+12} found that non-linear perturbations, such as the cosmological filament, are needed to explain the formation of cold clouds in the hot corona.

An alternative explanation is provided by galactic origin mechanisms that trigger the cooling of coronal gas. There is much observational evidence of a constant interaction between the cold gas in star-forming galaxies and their surrounding coronae. Sensitive \hi\ observations of star-forming disc galaxies reveal that $\sim 5-10 \%$ of \hi\ content of these galaxies is located a few kpc above the disc, forming the so-called \hi\ extragalactic layer \citep[e.g][]{Swaters97, Oosterloo+07}. In the Milky Way most of this emission consists of the Intermediate-Velocity Clouds (IVCs), cold gas complexes with disc-like metallicity located around $\sim 2$ kpc from the Sun \citep[e.g.][]{Marasco&Fraternali11}. Overall, the \hi\ halo kinematics is quite regular, but it rotates more slowly than the neutral component of the disc gas and its rotational velocity decreases with increasing height above the galactic plane. A vertical velocity gradient of $-15 \kms \rm{kpc^{-1}}$ has been measured in a number of spiral galaxies, including the Milky Way, both in the \hi\ neutral phase and in the ionized gas phase \citep[e.g.][]{Heald+06, Oosterloo+07, Marasco&Fraternali11}. Most of this gas is thought to be composed by the so-called `fountain clouds', gas ejected from the disc by stellar feedback \citep{Bregman80, Houck&Bregman90}, that travels through the coronal gas and eventually falls back to the disc in a time-scale of $\sim 80-100$ Myr. 

\citet{Fraternali&Binney06} built a dynamical model of fountain clouds that follow ballistic trajectories into the galactic halo and applied it to the extra-planar \hi\ observations of two nearby galaxies, NGC 891 and NGC 2403. This model is able to reproduce the vertical distribution of the \hi\ halo but underestimates the vertical velocity gradient: the observed rotational velocities are lower than the theoretical expectations. \citet{Fraternali&Binney08} found that the \hi\ halo kinematics could be explained by assuming that the fountain clouds lose angular momentum by accreting material with lower angular momentum from ambient medium. In order to investigate the physical phenomenon the drives the interaction between the hot corona and the cold fountain clouds, \citet{Marinacci+10, Marinacci+11} carried out a set of 2D hydrodynamical simulations of a cold (${T} = 10^4 \rm{K}$) and disc-like metallicity cloud traveling through the coronal halo of the Milky Way. They found that the cold fountain gas and the hot coronal gas mix efficiently in a turbulent wake behind the cloud and this mixing reduces dramatically the cooling time of the hot gas, triggering the condensation and the accretion of a fraction of the corona onto the disc. \citet{Marasco+12} used the dynamical model of \citet{Fraternali&Binney08} including condensation from the corona to reproduce the \hi\ halo of the Milky Way. They found a current accretion rate of coronal gas onto the disc of $\sim 2 \moyr$, in agreement with the accretion rate required by the Milky Way \citep[$\sim1-3\moyr$,][]{Chomiuk&Povich11,Putman+12}. \citet{Fraternali+13} further extended the model including warm gas in the turbulent wake generated by cloud-corona interaction. This new model is able to reproduce positions and velocities of the most of the warm absorbers observed in the Galactic halo. 

Accretion driven by galactic fountain seems to be a viable mechanism for star-forming galaxies to get the gas needed to sustain star formation. However, until now, it has been studied only under conditions representative of our Milky Way. It becomes crucial to extend the study of the cloud-corona interaction to different galactic environments. In this paper we extend the work done by \citet{Marinacci+10, Marinacci+11} investigating, through high-resolution hydrodynamical simulations, the cloud-corona interaction in haloes with different coronal temperatures, to probe a wide range of halo virial masses. These new simulations include the presence of thermal conduction, absent in the calculations performed so far. Thermal conduction may be a key process to determine the efficiency of the condensation. In fact, it may slow down or, under particular conditions, inhibit the condensation of coronal gas \citep[e.g.][]{Begelman&McKee90,Vieser&Hensler07}.

This paper is organized as follows. In Sec.\ref{Thermal Conduction} we illustrate the main features concerning thermal conduction and its hydrodynamycal treatment in the code. In Sec.\ref{Numerical simulations} we introduce the set of hydrodynamical simulations performed justifying the choices of the parameters. In Sec.\ref{Results} we present our simulations results focusing on the efficiency of coronal gas cooling and on the impact of thermal conduction on this process. In Sec.\ref{Discussion}, we discuss the limitations of these simulations and their implications for galaxy evolution and in Sec.\ref{Conclusions} we summarize our main results.

\section{Thermal Conduction}
\label{Thermal Conduction}

\subsection{Analytic theory}
\label{The classical theory}
Thermal conduction is the transfer of energy arising from a temperature gradient at the interface between two different gas phases. Hot electrons transfer heat to the colder medium and the net effect is a smoothing of the temperature gradient at the interface between the two fluids. According to the classical theory, thermal conduction in a fully ionized plasma is given by \citep{Spitzer62}:
\begin{equation}
\textbf{\textit{q}}_\mathrm{class} = - \kappa_\mathrm{Sp} \, \mathbf {\nabla} T \:,
\label{Spitzer}
\end{equation}
where \textbf{\textit{q}}$_\mathrm{class}$ is the so-called `heat conduction flux', ${\nabla} T$ is the temperature gradient, and the heat conduction coefficient is
\begin{equation}
\kappa_\mathrm{Sp} = \dfrac{1.84 \times 10^{-5} T^{5/2}}{\mathrm{ln\Psi}}  \: \: \: \mathrm{erg \,s^{-1} \,K^{-1}\, cm^{-1}} \:,
\label{Spitzer2}
\end{equation}
where $\ln\Psi$ is the Coulomb logarithm and it can be expressed as 
\begin{equation}
\mathrm{ln\Psi} = 29.7 + \mathrm{ln} \left[ \dfrac{T_{e} / 10^6 K}{\sqrt{n_{e} / cm ^{-3}}}\right]
\end{equation}
with $n_\mathrm{e}$ being the electron density and $T_\mathrm{e}$ the electron temperature.

In several astrophysical applications there are cases where the classical theory of thermal conduction is not directly applicable, and in particular the efficiency of thermal conduction is (strongly) reduced compared to the Spitzer value presented in Eq.~\ref{Spitzer}. For instance, in the presence of magnetic fields, the motion of the conducting electrons is not isotropic but parallel to the magnetic field lines, so the classical thermal conduction (Eq.~\ref{Spitzer}) is strongly reduced in the transverse direction. To take into account the effect of a tangled magnetic field, the Spitzer formula is usually multiplied by a dimensionless parameter \textit{f}, less than or of the order of unity:
\begin{equation}
\textbf{\textit{q}} = - f\, \kappa_\mathrm{Sp} \, \mathbf {\nabla} T \;.
\label{Spitzer+magnetic}
\end{equation}
\citet{Rechester&Rosenbluth78} and \citet {Chandran&Cowley98} estimated that, in the presence of a tangled magnetic field, the coefficient of thermal conduction is a factor of $\sim 100-1000$ lower than the Spitzer coefficient. \citet{Narayan&Medvedev01} found that if the turbulence extends on a wide range of length scales, as it might happen  with  strong-intermediate MHD turbulence, the efficiency of the Spitzer thermal conduction depends on the ratio between the minimum turbulence length scale and the relevant scale of the structure: \textit{f} increases with decreasing the minimum turbulence scale, converging around a maximum value of $\sim$ 0.2. 

The Spitzer formula also breaks down when the local temperature scale-length falls below the mean free path of the conducting electrons (the classical description is based on the assumption that the mean free path of the electrons is very short). In this case the heat flux is replaced by a flux-limited form the so-called `saturated heat flux' \citep{Cowie&McKee77}:
\begin{equation}
\vert \,\textbf{\textit{q}}_\mathrm{sat}\vert = 5 \Phi_{\mathrm{s}}\rho c^3\;,
\label{Saturated1}
\end{equation} 
with the sound speed \textit{c} and the density $\rho$. $ \Phi_{\mathrm{s}}$ is an efficiency factor less than or of the order of unity, which embodies some uncertainties connected with the flux-limited treatment and flux suppression due to magnetic fields. \citet{Dalton&Balbus93} introduced a formula that takes into account a smooth transition between two regimes:
\begin{equation}
\textbf{\textit{q}}_\mathrm{eff} = - \dfrac{\kappa_\mathrm{Sp} }{1+\sigma} \, \mathbf {\nabla} T\;,
\label{Saturated2}
\end{equation}
where $\sigma$ is the ratio between the classical heat flux and the saturated heat flux
\begin{equation}
\sigma = \dfrac{\kappa_\mathrm{Sp} || \nabla T||}{5 \Phi_{\mathrm{s}}\rho c^3}\;.
\label{Saturated3}
\end{equation}
$||\nabla T||$ being the magnitude of the local temperature gradient.
Equation \ref{Saturated2} guarantees that the thermal conduction is significantly reduced for $\sigma\geq 1$, and that the heat flux never exceeds the maximum saturated value.

\subsection{Hydrodynamical treatment}
\label{Hydrodynamical treatment}
The code used for our simulations is ATHENA \citep{Stone+08}, a grid-based, parallel and multidimensional magnetohydrodynamical code. The code implements algorithms based on higher-order Godunov methods, with a conservative finite-volume discretization to evolve volume averages of the mass, momentum, and total energy density.

We modified the module for isotropic thermal conduction present in ATHENA, using the analytic formula given by a combination of Eq. \ref{Spitzer+magnetic} and \ref{Saturated2}:
\begin{equation}
\textbf{\textit{q}} = - f\,\dfrac{\kappa_\mathrm{Sp} }{1+\sigma} \, \mathbf {\nabla} T .
\label{Eqcode}
\end{equation}
The effect of thermal conduction is taken into account in the energy equation: 
\begin{equation}
\dfrac{\partial e}{\partial t} + \mathbf {\nabla} \cdot [ (e+P) \mathbf{v}]= \,-\, \rho^2 \Lambda - \mathbf {\nabla} \cdot \textbf{\textit{q}}
\label{Energy}
\end{equation}
where \textit{e} is the energy density, \textbf{v} the velocity, \textit{$\rho$} the density and $P=(\gamma-1) U$ the pressure with $U$ the internal energy density and $\gamma=5/3$. \textit{$\Lambda$} is the radiative cooling rate as a function of temperature and metallicity. We used the tabulated values of \citet{Sutherland&Dopita93} in case of collisional ionization equilibrium. In ATHENA the cooling term is added to the energy equation at first-order via operator splitting and explicit treatment of the integration time. In our simulations at high resolution the cooling time is usually larger than the time-scale associated to hydrodynamic processes. However, to avoid possible problems of stability in regions where cooling is very effective, we limited the hydrodynamic time step to be a fraction ($10\%$) of the cooling time.
Thermal conduction is also directly added to the energy flux. Since the explicit update, present in ATHENA, involves very restrictive CFL constraint on the hydrodynamic time step, we implemented an implicit update for the temperature evolution. A detailed description of the implementation as well as a numerical test of the thermal conduction algorithm can be found in Appendix \ref{appendix}.

\subsubsection{Effects of magnetic fields}

In our simulations we did not include magnetic fields. In order to take into account their impact on the thermal conduction efficiency we used the \textit{f} factor (see Sec. \ref{Thermal Conduction} and Eq. \ref{Spitzer+magnetic},\ref{Eqcode}).

According to \citet{Narayan&Medvedev01} the maximum value of \textit{f} is 0.2. However their analysis is performed on a single plasma phase while our work focuses on the interaction between two different fluids, cloud and corona, with different magnetic fields. At the interface between two fluids with different magnetic fields, thermal conduction can be further reduced due to low efficiency of a phenomenon called `magnetic reconnection'. The magnetic reconnection breaks and reconnects magnetic field lines belonging to different fluids. When the process of reconnection is highly efficient, the conducting electrons are able to move and to transfer heat from the hot medium to the cold one. When the process of reconnection is weakly efficient, the conducting electrons tend to remain inside own fluid and thermal conduction is strongly reduced \citep[e.g.][]{Biskamp2000, Priest00}. In a chaotic magnetic field the role of the magnetic reconnection could be very hard to trace and certainly this is far from our current purposes. In order to take into account a possible thermal conduction reduction due to a less efficient magnetic reconnection we fix \,$\textit{f} =$ 0.1. We judge this to be a good compromise between the upper limit found by \citet{Narayan&Medvedev01} and a possible effect due to magnetic reconnection at the cloud-corona interface. In Sec. \ref{Limitations of our results} we also discuss a test with $f=0.2$.

Saturated thermal conduction may be also reduced by the presence of magnetic fields. We accounted for this effect through the efficiency factor $\Phi_{\mathrm{s}}$, where $\Phi_{\mathrm{s}} \sim \sqrt{f}$ \citep[see also][]{Cowie&McKee77}. This dependence can be understood through the following argument. If we call $\theta$ the angle between the local direction of the magnetic field and the local direction of the temperature gradient, the classical thermal conduction is reduced by a factor cos$^2 \theta$, where the first cos $\theta$ corresponds to the projection of the temperature gradient onto the magnetic field direction and the second cos $ \theta$ corresponds to the projection of the resulting heat flux, parallel to the magnetic field, onto the direction of the flux gradient. However, the saturated heat flux (eq. \ref{Saturated1}) does not depend on the temperature gradient, therefore it is only reduced by a factor cos $\theta$. In our simulations we assumed $\Phi_{\mathrm{s}} = \sqrt{f}$, neglecting other possible reductions of saturated thermal conduction besides suppression due to magnetic fields.

\section{Numerical simulations}
\label{Numerical simulations}

As mentioned in Sec.~\ref{Hydrodynamical treatment}, all our simulations were performed with the ATHENA code in a two-dimensional cartesian geometry. We performed just one three-dimensional simulation (Sim. 11 in Tab. \ref{ChangedParameters}), discussed in Sec. \ref{Limitations of our results}, in order to estimate the differences between the two geometries. The boundary conditions at four sides (six in case of a 3D geometry) are semi-permeable to allow for an outflow of gas from the computational domain.

As in \citet{Marinacci+10, Marinacci+11}, our simulations model a cold and metal-rich cloud that travels through a hot and static coronal gas with a given initial velocity (the ejection velocity from the galactic disc). The fountain clouds orbit over the disc in the gravitational potential of the galaxy and falls back onto the disc. However, in our simulations both gravitational acceleration and the coronal density variation along the cloud trajectory were neglected. This assumption is justified because, during their own orbit, the clouds do not change significantly their own distances from the galactic disc. In particular, they reach at most heights of few kiloparsec above the galactic plane and their distances from the Galaxy center vary by less than 30\% \citep{Fraternali&Binney06, Marasco+12}. Therefore, if the corona is in hydrostatic equilibrium with the gravitational potential of the galaxy, the coronal density is not expected to vary much.

\begin{table}
\centering
\begin{tabular}{cccccc}
\hline 
\hline
$T_\mathrm{cl}$& $v_\mathrm{ej}$& $n_\mathrm{cor}$& $Z_\mathrm{cor}$& $Z_\mathrm{cl}$\\
(K) & (km/s) & (cm$^{-3}$) & (Z$_\mathrm{\odot}$) & (Z$_\mathrm{\odot}$)\\
\hline
$10^4$&$75$ & $10^{-3}$&$0.1$ &$1.0$\\
\hline
\hline
\end{tabular} 
\caption{Initial parameters of all our simulations: ejection velocity $v_\mathrm{ej}$, coronal density $n_\mathrm{cor}$, cloud temperature $T_\mathrm{cl}$, coronal metallicity $Z_\mathrm{cor}$ and cloud metallicity $Z_\mathrm{cl}$. The cloud density is calculated by assuming pressure equilibrium between the cloud and the corona.}
\label{FixedParameters}
\end{table}

The parameters fixed in each simulation are listed in Table \ref{FixedParameters}. In all simulations the initial cloud temperature is $10^4$ K while the initial cloud velocity is $75 \kms$. This value represents the relative velocity threshold between the cloud and the ambient medium, below which the corona stops absorbing momentum \citep{Marinacci+11}. The coronal particle density is $10^{-3}$ cm$^{-3}$. This value is lower than the electronic density $n_\mathrm{e} = 2.6 \times 10^{-3}$ cm$^{-3}$ at $z = 10$ kpc found by \citet{Fukugita&Peebles06} and may be compared with the total particle density $n = 10^{-4}$ cm$^{-3}$ at $10$ kpc above the plane adopted by \citet{Heitsch&Putman09} or with the average electron density $<n_\mathrm{e}>= 5 \times 10^{-4}$ cm$^{-3}$ between the disc and $50$ kpc above it obtained by \citet{Anderson&Bregman11} through indirect evidence. The cloud metallicity is 1.0 Z$_\mathrm{\odot}$ while the coronal metallicity is 0.1 Z$_\mathrm{\odot}$, in agreement with the values estimated for the galaxies in which the hot halo was actually observed in the X-rays \citep{Bogdan+13, Hodges-Kluck&Bregman13, Anderson+16}. For the Milky Way the value is not well defined but studies through Far Ultraviolet absorption spectra and emission lines of \ovii\ and \oviii\ return values between $0.1$ and $0.3$ Z$_\mathrm{\odot}$  \citep{Sembach+03, Miller&Bregman15}. 

\begin{table}
\centering
\begin{tabular}{cccccccc}
\hline 
\hline
Sim.&$T_\mathrm{cor}$& $M_\mathrm{cl}$& $R_\mathrm{cl}$& Res.& TC& f & Geom.\\ 
&(K) & ($\mo$)& (pc)& (pc) & &\\
\hline
1&$2 \times 10^6$&$2.4 \times 10^4$&$100$&$2$ & OFF &- &2D\\
2&$2 \times 10^6$&$2.4 \times 10^4$&$100$&$2$ & ON &0.1&2D\\
3&$2 \times 10^6$&$2.4 \times 10^4$&$100$&$4$ & ON &0.1&2D\\
4&$2 \times 10^6$&$2.4 \times 10^4$&$100$&$1$ & ON &0.1&2D\\
5&$1 \times 10^6$&$1.2 \times 10^4$&$100$&$2$ & ON &0.1&2D\\
6&$4 \times 10^6$&$4.8 \times 10^4$&$100$&$2$ & ON &0.1&2D\\
7&$8 \times 10^6$&$9.6 \times 10^4$&$100$&$2$ & ON &0.1&2D\\
8&$8 \times 10^6$&$2.4 \times 10^4$&$60$&$2$ & ON &0.1&2D\\
9&$2 \times 10^6$&$2.4 \times 10^4$&$100$&$2$ & ON &0.2&2D\\
10&$2 \times 10^6$&$2.4 \times 10^4$&$100$&$10$ & ON &0.1&2D\\
11&$2 \times 10^6$&$2.4 \times 10^4$&$100$&$10$ & ON &0.1&3D\\
\hline
\hline
\end{tabular} 
\caption{List of the performed simulations. We varied both the coronal temperature $T_\mathrm{cor}$ (Sim. 2, 5, 6, 7) and the grid resolution (Sim. 2, 3, 4). The initial cloud radius, $R_\mathrm{cl}$, is 100 pc in all simulations, expect one where it is 60 pc (Sim. 8). Thermal conduction (TC) is turned off just in one simulation (Sim. 1). In all others, it is turned on with an efficiency of $10\%$ ($f = 0.1$, see Sec.~\ref{The classical theory}), except one (Sim. 9) with an efficiency of 20$\%$ ($f = 0.2$). The simulations were performed in a 2D cartesian geometry, except one (Sim. 11) performed in a 3D geometry.}
\label{ChangedParameters}
\end{table}

\begin{figure*}
\includegraphics[width=\textwidth]{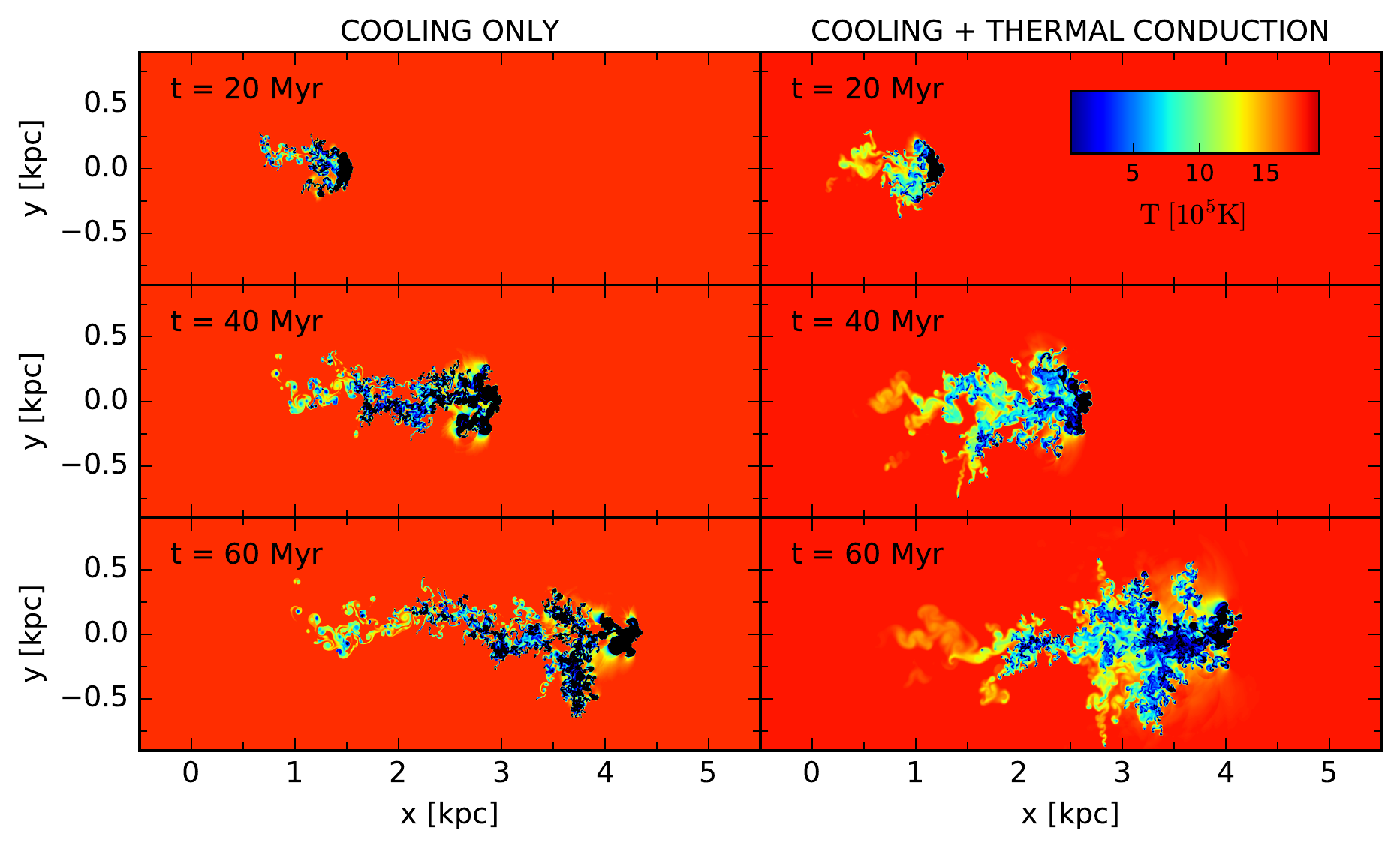}
\caption{Temperature snapshots of the simulations without (left panels) and with (right panels) thermal conduction with $T_\mathrm{cor}=2\times10^6$ K (Sim. 1 \& Sim. 2 in Tab. \ref{ChangedParameters}). The time at which the snapshots have been taken is indicated in each panel. The initial position of the cloud centre is x = 0 and y = 0.}
\label{Cond1}
\end{figure*}

In table \ref{ChangedParameters} the parameters that characterize the different simulations are listed: the coronal temperature, the grid resolution, the presence or absence of the thermal conduction and its efficiency, the cloud radius. As we said in Sec.~\ref{Introduction}, our main goal is to investigate the cloud-corona interaction and the efficiency of coronal gas condensation in environments with different temperatures. The coronal gas temperature is close to the virial temperature of the dark matter halo where the galaxy resides:
\begin{equation}
T_\mathrm{vir} \,=\, \frac{G \mu m_\mathrm{p} M_\mathrm{vir}}{2 k_\mathrm{B} r_\mathrm{vir}} \;,
\label{Tvir_relation}
\end{equation}
where $\mu$ is mean molecular weight of the coronal gas, $m_\mathrm{p}$ the proton mass, $k_\mathrm{B}$ the Boltzmann constant and $G$ the gravitational constant. In Eq. \ref{Tvir_relation}, $r_\mathrm{vir}$ is the virial radius, defined such that within a sphere of radius $r_\mathrm{vir}$, the average mass density of the galaxy halo is $\Delta_\mathrm{vir}$ times the critical density $\rho_\mathrm{cr}$ of the Universe:
\begin{equation}
r_\mathrm{vir} \equiv  \left( \dfrac{3 M_\mathrm{vir}}{{4 \pi}\, \Delta_\mathrm{vir}\,\rho_\mathrm{cr}} \right) ^{{1}/{3}}\,.
\label{Mvir_Rvir}
\end{equation}
$\Delta_\mathrm{vir}$ is the virial overdensity provided by the dissipationless spherical top-hat collapse, it is a function of the cosmological models, and it may vary with time. For the family of flat cosmologies ($\Omega_\mathrm{m}+\Omega_\mathrm{\Lambda}=1$), $\Delta_\mathrm{vir}$ can be approximated by $\Delta_\mathrm{vir} \simeq 18 \pi^2 + 82x -39x^2/\Omega_\mathrm{m}(z)$, where $x\equiv\Omega_\mathrm{m}(z)-1$ and $\Omega_\mathrm{m}(z)$ is the ratio of mean matter density to critical density at redshift \textit{z} \citep{Bryan&Norman98}. In the $\Lambda$CDM cosmological model with $\Omega_\mathrm{m}=0.3$ the value of $\Delta_\mathrm{vir}$ is $\sim 350$ at $z=0$.
In conclusion, the coronal temperature is a direct measure of the virial mass of the dark matter halo of the galaxy ($T_\mathrm{cor}\propto M_\mathrm{vir}^{2/3}$). Therefore, by changing the coronal temperature in our simulations we are exploring a wide range of virial masses.

We assumed pressure equilibrium between the cloud and the external medium at the beginning of each simulation. Indeed, we made experiments with out-of-equilibrium initial conditions and we found that the cloud readjusts itself and it reaches pressure equilibrium in $\sim 5$ Myr. In this way, the cloud number density is fixed by environmental parameters and it varies in simulations with different coronal temperature: higher coronal temperature means higher cloud density. The initial cloud radius is the same used by \citet{Marinacci+10}: it is set at a value of $100$ pc in all simulations except in Sim. 8. Then, since the initial radius is fixed, the mass of the cloud changes with the number density between the different setups. In particular, it ranges between $1.2 \times 10^4$ $\mo$ and $9.6 \times 10^4$ $\mo$ (see Sim. 2, 5, 6, 7 in Tab. \ref{ChangedParameters}). These cloud masses are consistent with the estimated masses of the Galactic IVCs for which good distance constraints exist \citep{Wakker01}. The Jeans mass of a standard cloud ($T_\mathrm{cl}=10^4$ K, $r_\mathrm{cl}=100$ pc) is $1.6 \times 10^8$ $\mo$, and so in our simulations the absence of self-gravity is justified.

The hydrodynamical simulations performed by \citet{Marinacci+10, Marinacci+11} started with unrealistically spherical clouds. Hydrodynamical simulations of supernovae explosions have indeed shown that the geometry of the gas ejected from the disc is strongly irregular \citep[e.g.][]{Melioli+13}. A non-spherical geometry implies a wider contact surface between cloud and corona and, then, a faster and more efficient interaction between them. In order to allow a quick deformation of the spherical cloud and to make the simulations more realistic, we introduced the presence of turbulent motions inside the cloud. We initialized the x-velocity and y-velocity of each cell comprising of the cloud by randomly sampling a gaussian distribution with a given velocity dispersion. For this quantity we used a value of $10 \kms$, in agreement the velocity dispersion, mainly due to turbulence, observed for the \hi\ in our Galaxy and in nearby disc galaxies \citep{Tamburro+09}. To impart a net initial velocity to the cloud, as specified in Table \ref{FixedParameters}, we added a bulk velocity of $75 \kms$ to the turbulent velocity field in the x-direction (the cloud's direction of motion).

Since we used a 2D geometry, one of the dimensions perpendicular to the cloud velocity has been suppressed and we are simulating flow around an infinite cylindrical cloud that is moving perpendicular to its long axis. The cylinder initially has a circular cross-section of radius $R_\mathrm{cl}$. From the simulations we obtained quantities per unit length of the cylinder and we related these to the corresponding quantities for an initially spherical cloud of radius $R_\mathrm{cl}$ by multiplying the cylindrical results by the length $4R_\mathrm{cl}/3$ within which the mass of the cylinder equals the mass of the spherical cloud. We calculated the values of the cloud mass in Tab. \ref{ChangedParameters}
 by using this correction.
 
\section{Results}
\label{Results}
Below we describe the main results of our hydrodynamical simulations, focusing on the mass transfer between the cloud and the ambient medium and on the growth of cold gas mass with time. We define `cold gas' all gas at temperatures below $10^{4.3}$ K. At temperature above $10^{4.3}$ K, the Hydrogen is almost completely ionized and the fraction of \hi\ is less than $10\%$ \citep[e.g,][]{Sutherland&Dopita93}.

\subsection{Thermal conduction effects at $\mathbf{T_\mathrm{cor}=2\times10^6 \, K}$}
\label{Thermal conduction effects (I)}

\begin{figure}
\includegraphics[width=0.48\textwidth]{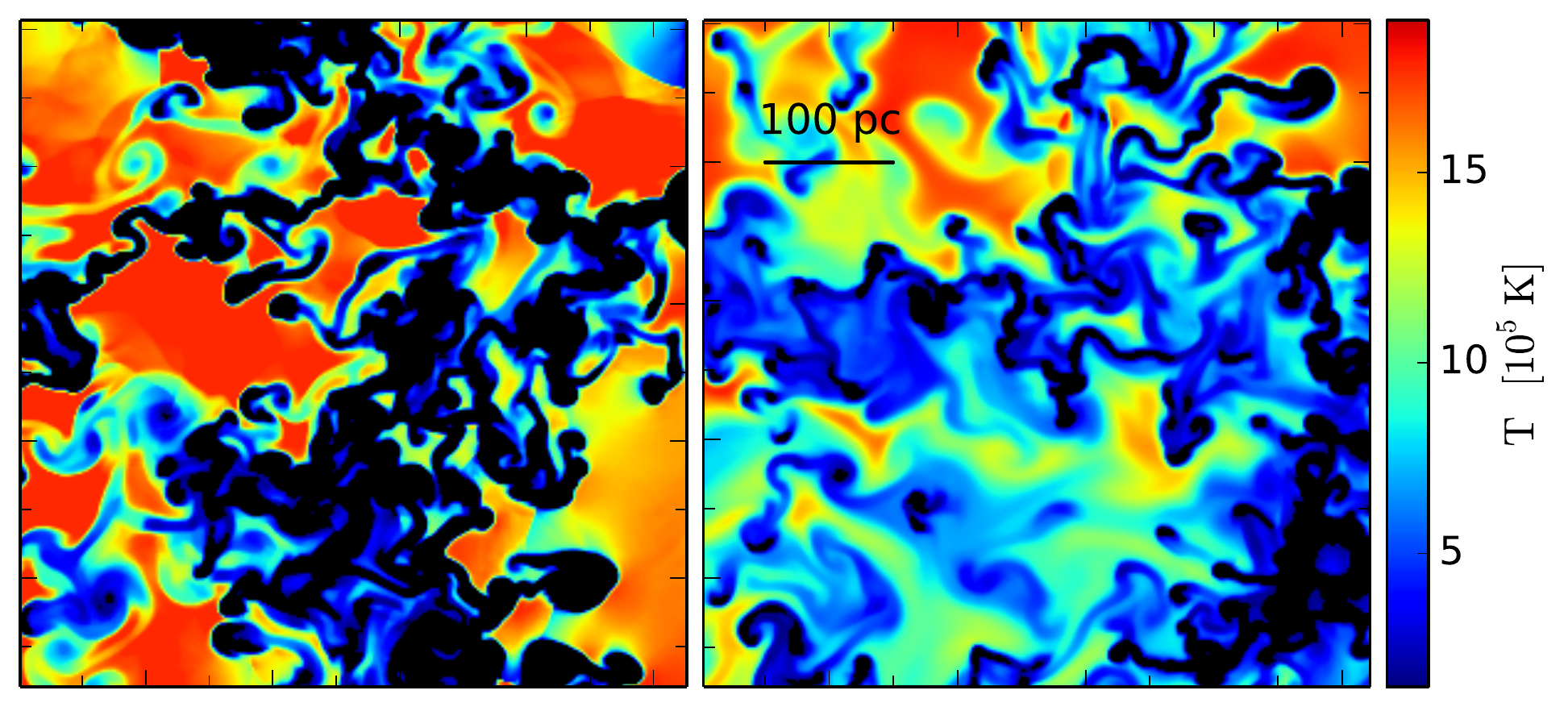}
\caption{Temperature map zoom of a portion of the wake behind the cloud of the simulations with $T_\mathrm{cor}=2\times10^6$ K without (left panel) and with (right panel) thermal conduction after 60 Myr.}
\label{Cond2}
\end{figure}

Fig. \ref{Cond1} shows the temperature distribution on our grid after 20, 40 and 60 Myr of the simulations without (left panels) and with (right panels) thermal conduction with $T_\mathrm{cor}=2\times10^6$ K (Sim. 1 and Sim. 2 in Tab. \ref{ChangedParameters}). Qualitatively, the evolution of the cloud in the present simulations is similar to that observed in the simulations of \citet{Marinacci+10,Marinacci+11}. Due to the ram pressure stripping arising from the motion, the cloud feels a drag which causes it to decelerate and to warp. Moreover, the relative motion between cloud and corona produces Kelvin-Helmholtz instabilities at the interface between the two fluids. The cold gas, stripped from the cloud, mixes efficiently with the hot coronal gas in a turbulent wake behind the cloud, where radiative cooling could become effective. 

The general evolution of the cloud appears to be quite similar: the two fluids mix creating turbulent wakes behind the cloud. However, by looking at Fig.~\ref{Cond1}, we can note some difference in the shape of the wake, which is less elongated but more laterally extended when thermal conduction is included. Fig.~\ref{Cond2} shows a magnification of a portion of the wake for both simulations.  In the case without thermal conduction (left panel) the wake is composed by numerous cloudlets in which gas is cooling down ($T \lesssim 5\times10^5$ K ), its general structure looks very compact. In the case with thermal conduction (right panel) the wake is composed by a mixture of gas at different temperatures: cold clouds and filaments are embedded in a hotter gas at temperature close to $10^6$ K. 

Thermal conduction is a diffusive process. Its effect is to create a more widespread and warmer wake in which the temperature gradients, due to the presence of colder structures, tend to be smoothed. Therefore, thermal conduction partially hinders the formation and survival of cold cloudlets, consequence of radiative cooling of the wake.

\subsection{Condensation of coronal material}
\label{Condensation of coronal material}

\begin{figure}
\includegraphics[width=0.48\textwidth]{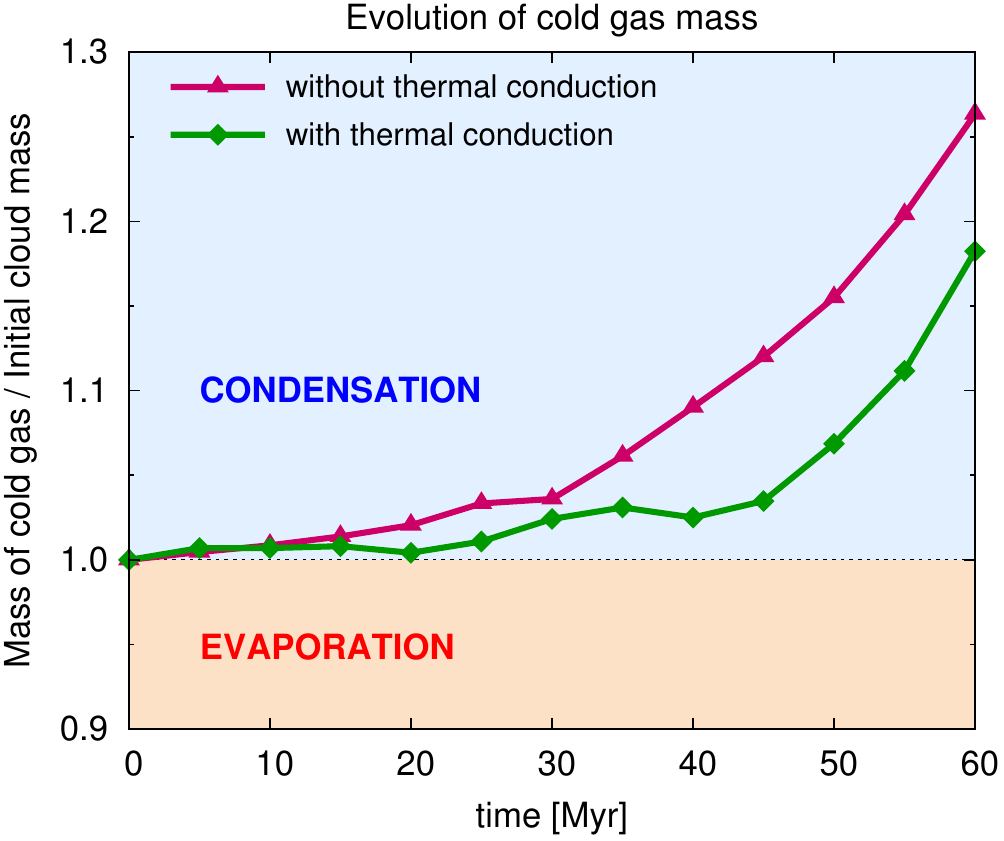}
\caption{Evolution of mass of cold gas ($T <10^{4.3}$ K) with time for two simulations with coronal temperature $2\times10^6$ K : one without thermal conduction and one with thermal conduction (Sim. 1 and Sim. 2 in Tab. \ref{ChangedParameters}).}
\label{Cond3}
\end{figure} 

The purpose of this work is to understand under what conditions cold clouds survive and cool down coronal gas during their motion or if they evaporate in the surrounding coronal medium. Thermal conduction can play an important role on gas evaporation or condensation. In Sec. \ref{Thermal conduction effects (I)}, we saw that thermal conduction tends to smooth the temperature gradients at the interface between different fluids, limiting the efficiency of radiative cooling and preventing the survival of the cold gas.

An analytic criterion to establish if radiative processes dominate thermal conduction was found by \citet{Begelman&McKee90}. They investigated the evolution of a two-phase medium consisting of clouds embedded in a hot plasma and showed that radiative cooling dominates if the length scales of the relevant structures, $l$, exceed a critical length:
\begin{equation}
\lambda_\mathrm{Field} \equiv \sqrt{ \dfrac{\kappa_\mathrm{Sp} T_\mathrm{hot}}{n_\mathrm{cold}^2 \Lambda (T_\mathrm{cold})}} \; .
\label{Field}
\end{equation}
the so-called `Field length', in view of \citet{Field65} work on thermal instabilities in astrophysical plasmas, in which he demonstrated that thermal instabilities are suppressed on scales smaller than this characteristic length. In Eq. \ref{Field} $T_\mathrm{hot}$ is the temperature of the hot plasma and $n_\mathrm{cold}$ and $\Lambda (T_\mathrm{cold})$ respectively the numerical density and the cooling rate of the cold gas. The Field length is the maximum length scale on which heat energy transport is effective: for $l \ll \lambda_\mathrm{Field}$ the temperature distribution of structures that are embedded in the hot plasma is dominated by the thermal conduction, while for $l \gg \lambda_\mathrm{Field}$ thermal conduction is negligible and the temperature evolution is dominated by radiative cooling.

In the simulations showed in Sec. \ref{Thermal conduction effects (I)}, $T_\mathrm{cor}=2\times10^6$ K and $n_\mathrm{cl}=0.2$ cm$^{-3}$, then the Field length is $\sim$ 20 pc\footnote{We used $\Lambda (T_\mathrm{cold}) = 10^{-23.93}$ ergs cm$^3$ s$^{-1}$, that is the average cooling rate in the temperature range between $10^4$ K and $10^{4.3}$ K, according to \citet{Sutherland&Dopita93}.} . This value is $\sim$ five times smaller than the initial cloud radius (100 pc), therefore a global evolution of the cloud quite similar in the presence or in the absence of thermal conduction is expected. However, the Field length is comparable with the size of cold cloudlets created in the wake (see Fig. \ref{Cond2}). At this point the fundamental issue is to understand whether these structures evaporate in the mixture or not because the growth of cold gas mass is determined by their survival.

We studied the evolution of mass of the cold gas in the two previous simulations, by extracting the mass of gas at temperature below $10^{4.3}$ K at different times. Fig. \ref{Cond3} shows the quantitative result. Both in the presence and in the absence of thermal conduction the amount of condensation increases with time, indicating that more and more coronal gas cools down in the wake. The mass profiles become nearly exponential after $\sim 30$ Myr in the absence of thermal conduction and after $\sim 40$ Myr in the presence of it. We interpreted this delay as the time required for cold gas, stripped from the cloud, to mix efficiently with the coronal gas plus the time required for the coronal gas to cool to $ T \lesssim 10^{4.3}$ K \citep[see also][]{Marasco+12}. The latter is longer in the presence of thermal conduction because the process of condensation is hindered. For this reason, the amount of condensation in the absence of thermal conduction is systematically larger: after 60 Myr the amount of cold mass is $\sim 26 \%$ of the initial mass cloud against $\sim 18 \%$ in the presence of thermal conduction. 

We can conclude that thermal conduction slows down the coronal gas condensation but does not inhibit it. Inside the warm wake at temperatures close to $10^6$ K, condensation on smaller scales occurs creating cold clouds and filaments. A fraction of these structures is able to survive for a time longer than the simulation time (60 Myr). This results has been obtained with a coronal temperature of $2\times10^6$ K, obviously the situation could change for different coronal temperatures, as we show in Sec. \ref{Thermal conduction effects (II)} and \ref{Condensation at different coronal temperatures}.

\begin{figure}
\includegraphics[width=0.48\textwidth]{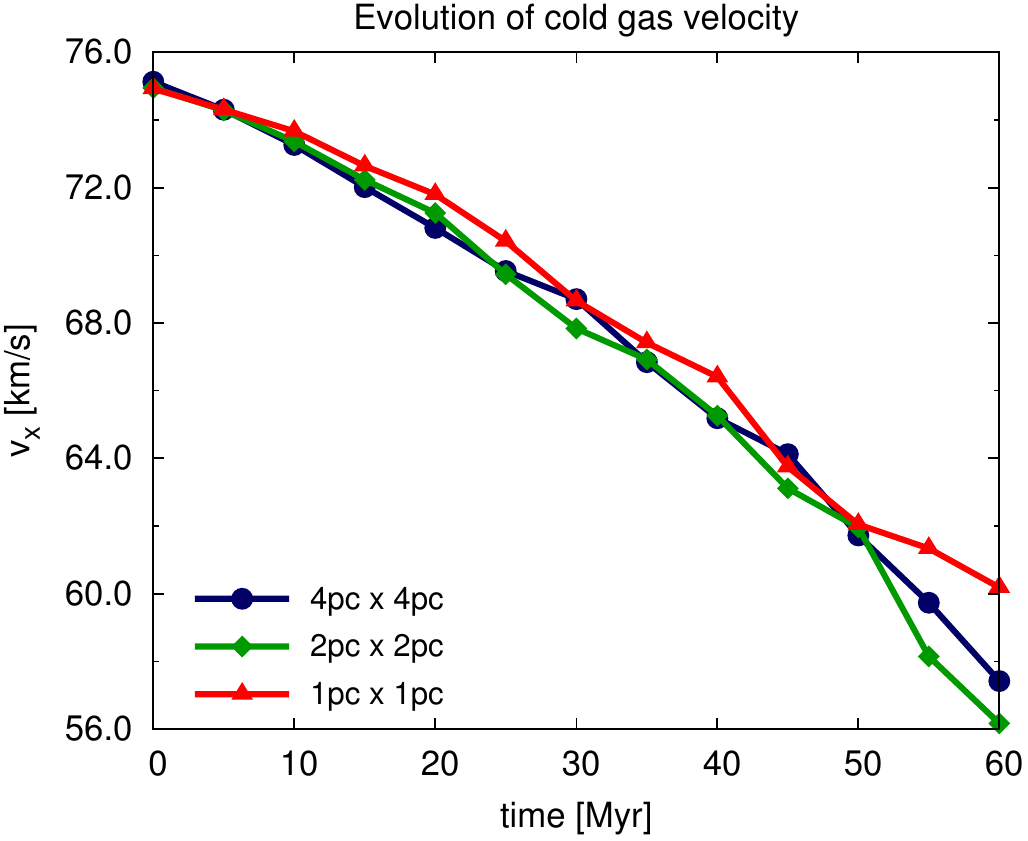}
\\
\\
\includegraphics[width=0.48\textwidth]{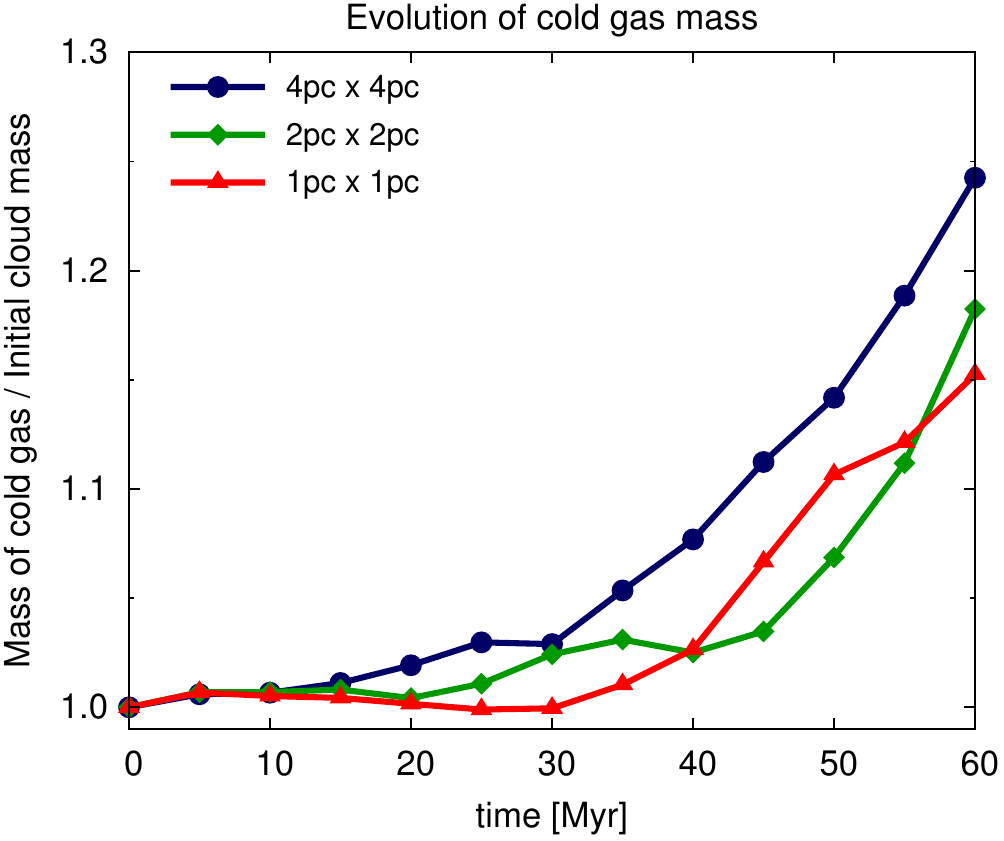}
\caption{Evolution of velocity (upper panel) and of mass (lower panel) of cold gas ($T <10^{4.3}$ K) with time for three different resolutions when the thermal conduction is switched on (Sim. 2, 3, 4 in Tab. \ref{ChangedParameters}).}
\label{Res1}
\end{figure} 

\subsection{Resolution}

We study the convergence of the code at different grid spacing by carrying out three simulations with different resolutions: 4pcx4pc, 2pcx2pc, 1pcx1pc (see Sim 2, 3, 4 in Tab. \ref{ChangedParameters}). In Fig.~\ref{Res1} we compare the results obtained from these three simulations. The upper panel shows the temporal evolution of the cold gas velocity centroid, defined
as the total momentum of the cold gas in the cloud direction of motion divided by the total mass of cold gas. There is a good agreement between all three trends, it seems that even the lower resolution is able to model the slowdown effect of the cloud due both to coronal ram-pressure stripping and to condensation of material that was originally at rest  \citep[see][for details]{Marinacci+10, Marasco+12}.

A less evident agreement is found in the lower panel of Fig.~\ref{Res1}, where we show the mass evolution of cold gas with time. The amount of cold gas in the simulation at lower resolution is systematically higher, while the simulations with intermediate and high resolution have roughly similar mass profiles intersecting one another in several points across the time and showing a difference not larger than a few percent. Different profiles at different resolutions are related to the evolution of material stripped from the cloud and to its mixing with the coronal gas. As we saw in Fig.\ref{Cond1} and Fig. \ref{Cond2}, the wake of the cloud is composed by small cloudlets and filaments. In a numerical simulation, numerical diffusion truncates this hierarchy of substructures on a scale of a few times the grid resolution: the numerical diffusion tends to smooth temperature gradients so the higher the resolution of the wake, the greater the concentration of the cold material in substructures. Increased resolution inhibits the phase mixing and the subsequent coronal gas condensation. Therefore, in simulations with radiative cooling but without any physical diffusive processes (e.g. thermal conduction), it is in practice very difficult to reach convergence between results at different resolutions: the higher the resolution, the lower the amount of condensation.

\begin{figure*}
\includegraphics[width=\textwidth]{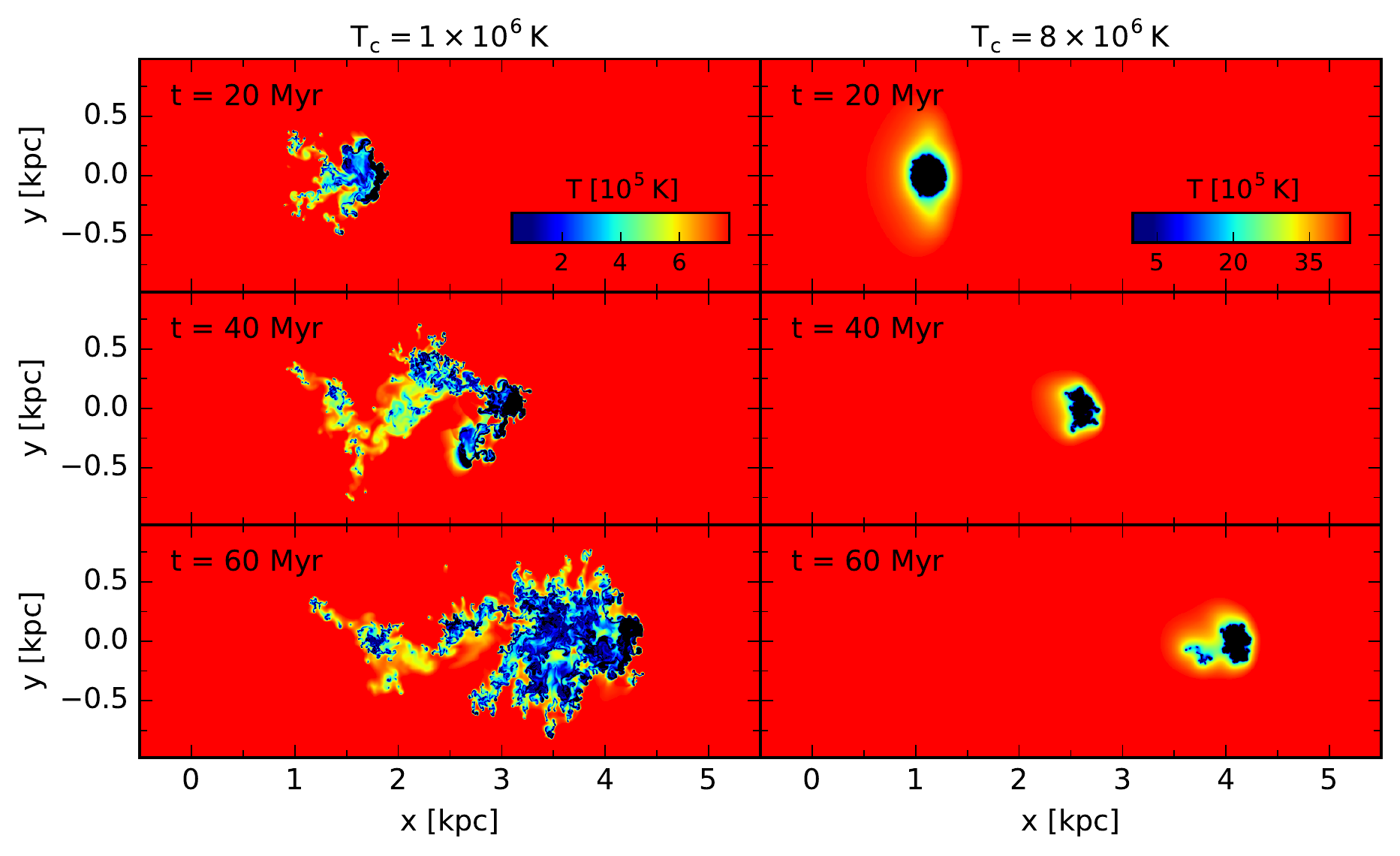}
\caption{Temperature snapshots of the simulations with $T_\mathrm{cor}=10^6$ K (left panels) and with $T_\mathrm{cor}=8\times10^6$ K (right panels) in the presence of thermal conduction (Sim. 5 and Sim. 7 in Tab. \ref{ChangedParameters}). The time at which the snapshots have been taken is indicated in each panel. The initial position of the cloud centre is x = 0 and y = 0.}
\label{Tvir2}
\end{figure*}

It appears that the introduction of thermal conduction has been helpful to restrain the numerical diffusion. As we said in Sec. \ref{Thermal conduction effects (I)}, thermal conduction is a diffusive process: like numerical diffusion, it destroys the smaller structures smoothing the temperature gradients and creating a homogeneous mixture at intermediate temperature. However, unlike numerical diffusion, the scale at which thermal conduction operates does not depend on the grid resolution, but on physical conditions of the problem, in particular on the temperature of the hot coronal gas (see Eq. \ref{Spitzer}). Thus, although at very low resolutions numerical diffusion is the dominant diffusive process, increasing the resolution, thermal conduction can become the dominant process. The development of turbulence in the wake of our fountain clouds produces a hierarchy of smaller and smaller structures, following a power law distribution. The size of the smallest structures is determined by resolution, however their survival is largely determined by thermal conduction which operates efficiently on scales below the Field length ($\sim$ 20 pc). Therefore, the smallest clouds (a few parsec) do not influence the growth of cold gas mass because they are not able to survive, mix with hotter gas and trigger condensation. The convergence in condensation shown in Fig.~\ref{Res1} may reflect this change of regime.

After these considerations we can conclude that the resolution 2pcx2pc is good enough for our kind of simulations and we proceed with this resolution for the rest of the paper.

\subsection{Thermal conduction effects at different $\mathbf{T_\mathrm{cor}}$}
\label{Thermal conduction effects (II)}

Thermal conduction strongly depends on the temperature of the hot gas. The heat conduction flux, \textit{\textbf{q}}, is proportional to $T^{5/2}$ (see Eq. \ref{Spitzer2}), then small differences of temperature lead to large differences of exchanged heat. The consequence is that at high temperatures cold structures can evaporate much faster. However, as we saw in Sec. \ref{Thermal Conduction}, when the temperature changes on scales smaller than the mean free path of the conducting electrons, thermal conduction enters in a regime of saturation and the correlation between high temperatures and high heat transfer can be mitigated. A criterion to ascertain the effects of saturated thermal conduction on a structure with length scale $ l $ is given by the global saturation parameter:
\begin{equation}
\sigma_\mathrm{0}={\left (\dfrac{T_\mathrm{cor}}{1.54\times10^7} \right )}^2 \dfrac{f}{\Phi_\mathrm{s} n_\mathrm{cor} l_\mathrm{[pc]}} \;.
\end{equation}
For $\sigma_\mathrm{0} \gtrsim 1$ the effect of saturated thermal conduction becomes significant \citep{Cowie&McKee77,Dalton&Balbus93}.
In our simulations, the effect of saturation is negligible for the global evolution of the cloud: $\sigma_\mathrm{0}  \ll1$ for $l = R_\mathrm{cl} = 100 $ pc, regardless of coronal temperature. However, the saturation effect on the cold cloudlets in the turbulent wake could become more and more important ($\sigma_\mathrm{0} \propto l^{-1}$) for increasing coronal temperature. Only at $T_\mathrm{cor}=8\times10^6$ K  $\sigma_\mathrm{0} \lesssim 1$ and at these high temperatures the saturation effect could have a slight influence on the general cloud evolution.

We investigated the evolution of cold clouds in coronae at different temperatures and we found that it is closely related to the importance of thermal conduction at a given temperature. Fig. \ref{Tvir2} shows the temperature distribution on the grid after 20, 40 and 60 Myr for the simulations with with $T_\mathrm{cor}=10^6$ K (left panels) and with $T_\mathrm{cor}=8\times10^6$ K (right panels) in the presence of thermal conduction (respectively Sim. 5 and Sim. 7 in Tab. \ref{ChangedParameters}). The general evolution of the cloud is very different in the two cases. In the simulation with $T_\mathrm{cor}=10^6$ K the situation is quite similar to the case analysed in Sec. \ref{Thermal conduction effects (I)}, with $T_\mathrm{cor}=2\times10^6$ K (see right panel in Fig. \ref{Cond1}). The cold gas, stripped from the cloud, mixes with the hot coronal gas in the turbulent wake behind the cloud. The wake is composed by large amount of gas at temperature below $2\times10^5$ K, indicating a very effective radiative cooling. Instead, in the simulation with $T_\mathrm{cor}=8\times10^6$ K the cloud nearly behaves like a rigid body. The turbulent wake does not exist, with the consequence that the mixing between the two fluids is absent.

It is important to point out that at $T_\mathrm{cor}=10^6$ K the Field length is $\sim$ 10 pc, while at $T_\mathrm{cor}=8\times10^6$ K it is $\sim$ 70 pc (see Eq.~\ref{Field}), a value comparable with the initial cloud radius (100 pc). Therefore, while at low temperatures thermal conduction affects only the cold cloudlets in the turbulent wake (see also Sec. \ref{Thermal conduction effects (I)}), at high temperature it can influence the global motion of the cloud. The effect of thermal conduction is to smooth the temperature gradient at the interface between the two fluids, creating a transition region at intermediate temperatures (this is clearly visible in the right panels in Fig. \ref{Tvir2}). 
The temperature smoothing leads the cloud to lose momentum at the contact surface. The resulting absence of a strong velocity gradient prevents the formation of hydrodynamical instabilities at the cloud-corona interface, involving very long destruction times for the cloud \citep[see also][]{Vieser&Hensler07b}. The main consequence of this phenomenon is the delay or the lack of the mixing phase between cloud and corona. 

In the latter case the initial cloud mass is $9.6\times10^4\,\mo$, the largest value that we explored in our simulations. In addition to thermal conduction, an inefficient coronal ram pressure could contribute to the slow destruction of massive clouds: large masses imply a large drag time \citep[$t_\mathrm{drag}\propto M_\mathrm{cl}$,][]{Fraternali&Binney06} and, then, a less efficient coronal ram pressure. In order to understand which of these two phenomena dominate at $T_\mathrm{cor}=8\times10^6$ K, we performed a simulation with a cloud four times less massive and a radius almost two times smaller, $M_\mathrm{cl}=2.4\times10^4\,\mo$ and $R_\mathrm{cl}=60$ pc (see Sim. 8 in Tab. \ref{ChangedParameters}). We found that the cloud evolution does not change significantly with respect to the standard simulation with a larger and more massive cloud: the cold gas remains in a compact configuration at the cloud head and the turbulent wake is absent. Then, we conclude that the strong effect of thermal conduction at $T_\mathrm{cor}=8\times10^6$ K drives the cloud evolution, regardless of its own initial mass and size.

\subsection{Condensation at different $T_\mathrm{cor}$}
\label{Condensation at different coronal temperatures}

\begin{figure}
\includegraphics[width=0.48\textwidth]{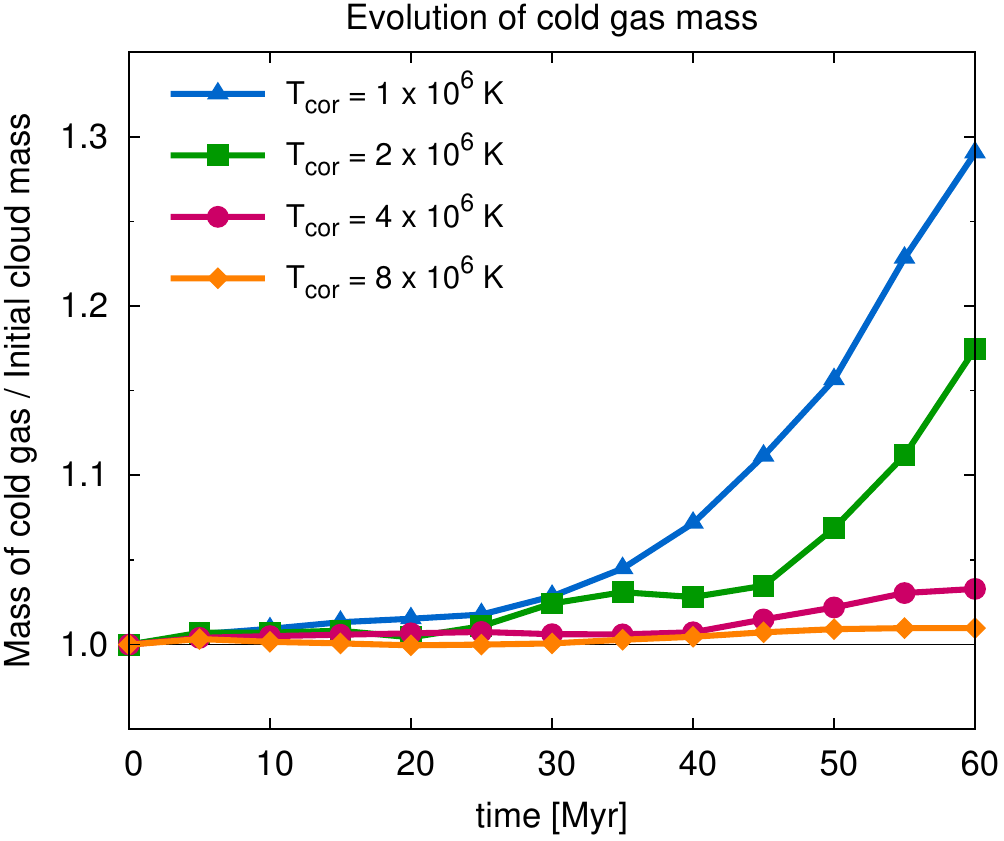}
\caption{Evolution of mass of cold gas ($T <10^{4.3}$ K) with time for four different values of coronal temperature: $1\times10^6, 2\times 10^6, 4\times 10^6$ and $ 8 \times 10^6$ K (respectively Sim. 5, 2, 6, 7 in Tab. \ref{ChangedParameters}).}
\label{Tvir}
\end{figure}

In order to understand how the galactic environment influences the coronal gas condensation, we analysed the evolution of mass of cold gas with time in four simulations with different coronal temperatures, $1\times10^6, 2\times 10^6, 4\times 10^6$ and $ 8 \times 10^6$ K (respectively Sim. 5, 2, 6 and 7 in Tab. \ref{ChangedParameters}). Fig.~\ref{Tvir} shows the results in the presence of thermal conduction. The amount of condensation becomes less efficient for increasing coronal temperature. After 60 Myr the mass of condensed gas is $\sim 30\%$ of the initial mass of the cloud for $T_\mathrm{cor}=1\times10^6$ K, $\sim 18\%$ for $T_\mathrm{cor}=2\times10^6$ K,  $\sim 4\%$ for $T_\mathrm{cor}=4\times10^6$ K and less than $1\%$ for $T_\mathrm{cor}=8\times10^6$ K.

In our simulations, the cloud temperature is fixed ($T_\mathrm{cl}=10^4$ K), therefore, higher coronal temperature means higher mixture temperature and, then, longer cooling times. For this range of temperatures the shape of the cooling function is crucial \citep[see][]{Sutherland&Dopita93}. At $T\sim10^6$ K a rising or falling of the temperature could have a great impact on the cooling time. In coronae with temperatures much larger than $2\times10^6$ K, the turbulent wake does not reach temperature low enough to trigger an exponential increase of condensation.

Thermal conduction can also play an important role in the different coronal gas condensation at different temperatures. The higher the coronal temperature, the more efficient the thermal conduction. As we show in Sec. \ref{Thermal conduction effects (I)} and \ref{Thermal conduction effects (II)}, thermal conduction has two important effects. It smooths the velocity gradient at the cloud-corona interface, making the cloud more compact and preventing the formation of hydrodynamical instabilities and subsequent mixing. This phenomenon causes a very slow cloud destruction (see Sec. \ref{Thermal conduction effects (II)}). The second effect is that once the gas is stripped from the cloud, thermal conduction changes its role, accelerating the heating of the cold gas and its evaporation in the coronal medium (see Sec~\ref{Thermal conduction effects (I)}).

We note that at $T_\mathrm{cor}=8\times10^6$ K the evolution of the cold mass is nearly flat. This means that the cold cloud does not acquire coronal gas and, at the same time, it does not lose its own mass. In this regime the cloud radius is comparable with the Field length (see Sec.~\ref{Thermal conduction effects (II)}), therefore the effect of radiative cooling is also important. Moreover, in Sim. 8 (see Tab.~\ref{ChangedParameters}), where the cloud radius is slightly smaller than the Field length, the equilibrium between thermal conduction and radiative cooling is evident. In this case the amount of cold mass decreases at a very slow rate: at the end of the simulation the cloud has lost $\sim 1 \%$ of its own initial mass. At higher temperature the radiative cooling is likely to become ineffective, leading the cloud evaporation in the coronal medium. 

\section{Discussion}
\label{Discussion}
\subsection{Limitations of our results}
\label{Limitations of our results}

We discuss physical and numerical issues that can affect our estimates about the growth of the cold gas mass during the simulation.

\begin{figure}
\includegraphics[width=0.48\textwidth]{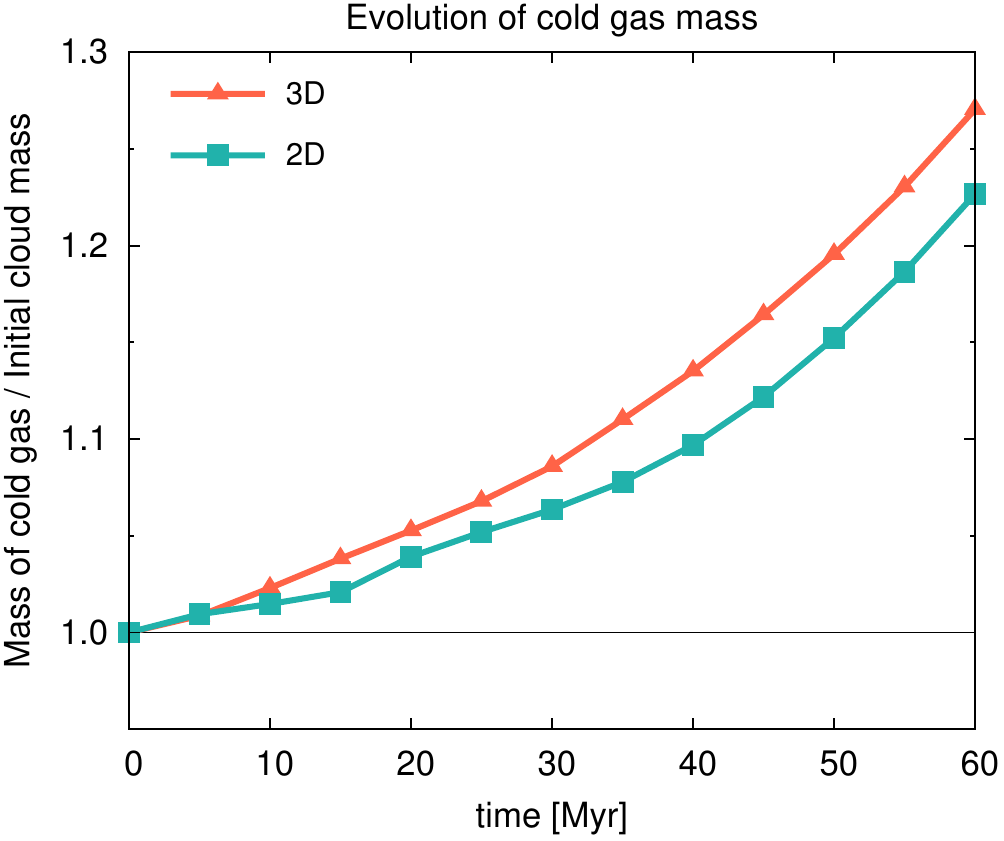}
\caption{Evolution of mass of cold gas ($T <10^{4.3}$ K) with time for two simulations, one in 3D and one in 2D cartesian geometry (Sim. 10 and Sim. 11 in Tab. \ref{ChangedParameters}).}
\label{3D2D}
\end{figure}

All our simulations were performed in a two-dimensional cartesian geometry. In order to estimate the limitations of the two-dimensional geometry, we performed one three-dimensional simulation at low resolution, 10pcx10pc (higher resolutions would have imply very long computational times for a code with static grid, as ATHENA). In Fig. \ref{3D2D} we compare the results obtained by using the two different geometries, and in particular the evolution of the mass of cold gas with time. The two simulations were performed with the same initial parameters and the same grid resolution (Sim. 10 and Sim. 11 in Tab. \ref{ChangedParameters}). The amount of cold gas increases with time in both simulations, it increases faster in the 3D case but the difference is not large: less than $15 \%$. After 60 Myr the fraction of condensed mass is $\sim 27 \%$ of the initial mass of the cloud in the 3D case, $\sim 23 \%$ in the 2D case. 
The nature of the cloud-corona interaction is different in 2D and 3D simulations. In 2D we simulate an infinite cylinder that is moving perpendicular to its long axis, while in 3D we actually simulate a spherical cloud. The result is that in 3D the contact surface between the two fluids is larger leading to additional developments of hydrodynamical instabilities and a more rapid phase mixing between the cold gas, stripped from the cloud, and the coronal gas. This result is in agreement with recent works that studied the differences between the two geometries through the evolution of a spherical cloud in a hotter and shocked medium \citep{Pittard&Parkin16}.
In conclusion, the 2D geometry slightly underestimates the growth of the cold gas mass.

Heating processes, like photoionization by background sources or by the star-forming disc, were neglected at this stage. Heating effect could be to render our estimates as upper limit, since it slows down the cooling gas process. At fixed metallicity radiative cooling rates depend mainly on the gas temperature, then, once the gas in the turbulent wake reaches values of temperature close to the peak of the cooling function, the process of condensation starts with an exponential trend (see Sec. \ref{Condensation of coronal material}). Heating rates strongly depend on the gas density, the lower the density the more efficient the heating process. Therefore, the gas density in the wake must become high enough in order that the cooling process begins efficiently. We expect that the main effect of heating is a delay in the time of start of the condensation.

The efficiency of magnetic field in suppressing thermal conduction, that we quantified through the factor \textit{f} (see Sec. \ref{Hydrodynamical treatment}), can also affect our findings. We remind that, according to \citet{Narayan&Medvedev01}, in the presence of a tangled magnetic field the efficiency of the Spitzer thermal conduction is reduced to 20$\%$ ($f=0.2$). However, for reasons that we explained in Sec. \ref{Hydrodynamical treatment}, we preferred to use $f=0.1$ in our simulations. We performed only a simulation with $f=0.2$ (Sim. 9 in Tab. \ref{ChangedParameters}). The parameters of this simulation are the same of the simulations discussed in Sec. \ref{Condensation of coronal material},  with $T_\mathrm{cor}=2\times10^6$ K. We found that after 60 Myr the amount of cold gas is $\sim 9\%$ of the initial mass of the cloud for $f=0.2$, while it is $\sim 18\%$ for $f=0.1$. In this case, the difference of a factor 2 in thermal conduction efficiency corresponds to a difference of a factor 2 in amount of condensation. The value of $\sim 9\%$ represents a lower limit of the condensation that we can obtain by varying the efficiency of thermal conduction through \textit{f}. The upper limit is instead given by the simulation in the absence of thermal conduction, that corresponds to assume $f=0$. In this case the amount of cold gas is $\sim 25\%$ of the initial mass. We conclude that in the interval of possible values of $f$ the process of condensation always occurs and the amount of accreting cold gas correlates with the efficiency of thermal conduction.

\subsection{Implication for the evolution of disc galaxies}
\label{Evolution of disc galaxies}
Star-forming galaxies like the Milky Way need some supply of external gas to feed star formation at the observed rate of $\sim 1 \moyr$ (see Sec. \ref{Introduction}). There are strong indications from cosmological models that the gas accreted by these galaxies to form stars must come from the intergalactic medium, which accumulates around galaxies in the form of hot corona. Hence, the evolution of a star-forming galaxy is strongly influenced by the flows of gas between the galaxy and its surrounding environment.

Our simulations show that the hot coronal gas can be efficiently cooled by the interaction with cold and metal-rich fountain gas at the disc-corona interface. Coronal gas condenses into a turbulent wake behind the cloud becoming fresh material that may sustain star formation at the current observed rates. However, the amount of coronal condensation strongly depends on the galactic environment as it becomes less efficient for increasing coronal temperature. The Milky Way coronal temperature is $\sim 2 \times 10^6$ K \citep{Fukugita&Peebles06, Miller&Bregman15}. In environments with this coronal temperature, the coronal gas condensation is efficient: after 60 Myr our cloud with initial mass $2.4\times10^4\mo$ condenses an amount of coronal gas equal to $\sim 18\%$ of its own mass (see Sec. \ref{Thermal conduction effects (I)} \& \ref{Thermal conduction effects (II)}).

Both model and observations show that the coronal temperature of a galaxy is roughly its virial temperature, which in turn is a measure of the virial mass of the dark matter halo where the galaxy resides. Using equations \ref{Tvir_relation} and \ref{Mvir_Rvir}, we related the coronal temperature of our simulations to the virial mass of the galaxy. Coronal temperatures of $10^6$ K correspond to virial masses of $\sim 10^{12} \,\mo$, coronal temperatures of $ 8\times10^6$ K correspond to a virial masses larger than $10^{13} \mo$. The virial mass corresponding to $2\times 10^6$ K is $\sim 3\times10^{12}\, \mo$. Actually, the virial mass estimated for the Milky Way \citep[$\sim 2\times10^{12} \, \mo$, e.g.][]{Li&White08} is slighly lower than this value. We note however that the observations of the Milky Way corona are quite sensitive to the coronal medium close to the disc where feedback may cause slight departures from the virial temperature \citep[e.g.][]{Strickland+04}. 

Our simulations have shown that the ability of fountain clouds to cool their surrounding corona is very efficient in galaxies with coronal temperature $\lesssim 2\times 10^6$ K / virial masses $\lesssim 3\times10^{12}\, \mo$, it is poorly efficient in galaxies with coronal temperature $\gtrsim 4\times10^6$ K /  virial masses $ \gtrsim 8 \times 10^{12}\,\mo$, and it is absent in galaxies with coronal temperature larger than $\gtrsim 8\times10^6$ K / virial masses $\gtrsim 10^{13}\,\mo$. Thus, there appear to be a mass threshold beyond which the cooling and accretion of the corona is not viable for galaxies ($3\times10^{12}\, \mo \lesssim M_\mathrm{vir} \lesssim 8\times 10^{12}\,\mo$).

\citet{Schawinski+14} showed that star-forming galaxies are mostly in haloes with low virial mass, while disc galaxies located in the green valley region, between the blue cloud of star-forming galaxies and the red sequence of quiescent galaxies in the colour-mass diagram, are almost exclusively galaxies with high virial mass, tipically larger than $10^{12} \mo/h$, where $h = H_\mathrm{0}/$100 km s$^{-1}$ Mpc$^{-1}$ is the Hubble parameter. Green valley galaxies are objects that are moving off or have already left the main sequence of star formation \citep[e.g.][]{Brinchmann+04, Peng+10, Schawinski+14}. Since all star-forming galaxies are on the main sequence and since green-valley galaxies must have experienced star formation in the past, by definition some process has turned off star formation but, to date, the ways this quenching of the star-formation occurs are still matter of debate. In terms of the Hubble fork, the disc galaxies with the lowest ratio between the current and the past SFRs are early-type disc galaxies, near the S0/Sa locus \citep[e.g.][]{Boselli+01}. Then, while late-type disc galaxies are undergoing a current phase of star formation, early-type disc galaxies have experienced a decline of star formation in response of exhaustion of gas reservoir. Milky Way galaxies, with virial mass close to the mass limit between the blue cloud and the green valley region, $M\gtrsim10^{12} \mo$, are still forming stars at slowly declining rates.
Since a relation between star formation history and virial mass of galaxy exists, it appears that environmental effects may regulate gas supply in galaxies. Many recent works showed that haloes below a critical shock-heating mass, $M_\mathrm{shock} \sim 2-3 \times 10^{11}\, \mo$, enjoy gas supply by cosmological cold streams (\textit{cold mode} accretion), not heated by a virial shock, and form stars, while the cold gas accretion is turned-off above this mass \citep[e.g.][]{Dekel&Birnboim06, keres+09}. According to these works, the shut off of gas supply prevents further star formation, leading the galaxies to evolve passively from the blue cloud to the red sequence as consequence of the depletion of the gas reservoir. However, as we said previously, Milky Way galaxies, with $M > M_\mathrm{shock}$, are still forming stars with a rate remained almost unchanged during their life time and mechanisms able to feed the star-formation are needed. 

The mechanism of coronal condensation driven by fountain clouds we proposed seems to suggest a suitable solution for explaining the presence of galaxies with $M > M_\mathrm{shock}$ on the main sequence of star formation. We note that our model does not exclude but it integrates the quenching mechanism due to the turning-off of \textit{cold mode} accretion. Our study shows that the coronal gas cools down efficiently in galaxies with low-intermediate virial mass but the ability of the galaxy to cool their corona decrease going from late-type to early-type disc galaxies, inhibiting accretion of cold gas available for star formation and leading potentially to the quenching of the galaxy. Moreover, given that the condensation is the consequence of mixing between the corona and the cold disc gas, it follows that early-type galaxies of any virial mass must be inefficient in cooling their corona because they have much less cold gas \citep[e.g.][]{Catinella+10}. 

Our results therefore suggest that the turn-off of the star formation in disc galaxies may be due not only to heating mechanisms of the hot halo, such as AGN \citep[e.g.][]{Ciotti+12} or stellar thermal feedback \citep[e.g.][]{Stinson+13} but also to the inability to cool their own coronal material experienced by massive galaxies ($M_\mathrm{vir}>3\times10^{12}\mo$). We can also speculate that our mechanism could explain both the quenching processes identified by \citet{Peng+10}: the stellar-mass and the environmental quenching. Indeed, the only effect that determines the star-formation quenching in our model is the ambient temperature. It correlates both with the stellar mass of galaxies (as mentioned), at least for isolated galaxies \citep{Behroozi+13,moster+13}, and with the kind of environment where galaxies live. For example, dense environments, as galaxy clusters, are characherized by high virial temperatures. In these ambients, not only the massive galaxies but also the late-type disc galaxies are not able to cool their surrounding corona, thus leading to the quenching of their star-formation.

\section{Summary and conclusions}
\label{Conclusions}

Cold gaseous discs of spiral galaxies are embedded in extended hot coronae of virial-temperature gas ($T\gtrsim10^6$ K) that may contain a significant fraction of so-called missing baryons. Milky Way-like galaxies might sustain their star formation at the current observed rates by transferring gas from the corona to the star-forming disc. In the region at the disc-corona interface there is a continuous interaction between the galaxy and its surrounding corona: cold fountain clouds ($T\sim10^4$ K), ejected from the disc by stellar feedback, travel through the hot coronal gas and interact with it. \citet{Marinacci+10, Marinacci+11}, through hydrodynamical simulations, studied the physical phenomenon the drives this interaction in an environment representative of our own Milky Way. They found that the cold fountain gas and the hot coronal gas mix efficiently and this mixing reduces dramatically the cooling time of the hot gas, triggering the condensation and the accretion of a fraction of the corona onto the disc. This new fresh coronal gas could become gas available for the star-formation.

In this paper we extended the work done by Marinacci et al. (2010, 2011) investigating, through high-resolution hydrodynamical simulations, the cloud-corona interaction in environments with different coronal temperatures. The new simulations were performed in the presence of radiative cooling and isotropic thermal conduction, the latter absent in the previous works. From these simulations we can draw the following conclusions.

\begin{itemize}
\item At $T_\mathrm{cor}\lesssim 2\times10^6$ K, formation of hydrodynamical instabilities at the interface cloud-corona triggers the loss of cold gas from the cloud. This gas mixes efficiently with the hot coronal material in a turbulent wake, where radiative cooling acts by forming cold gas cloudlets. The effect of thermal conduction is to prevent the survival of these cold structures by creating a warmer wake with smoothed temperature gradients. Nevertheless, thermal conduction is not able to inhibit the coronal gas condensation: after $\sim 30-40$ Myr, the latter starts with a trend nearly exponential and, after 60 Myr, the amount of cold gas is $\gtrsim 18\%$ of the initial cloud mass.
\item Increasing the coronal temperature ($T_\mathrm{cor}\gtrsim{4\times10^6}$ K), the effect of thermal conduction is to smooth the velocity gradient at the interface between the cloud and corona, preventing the formation of hydrodynamical instabilities and the subsequent cloud destruction. This phenomenon makes the cloud more compact and hinders the mixing phase between the two fluids. 
\item The amount of coronal condensation strongly depends on the environment temperature as it becomes less efficient for increasing coronal temperature. This trend is due to the combined effect of radiative cooling, that dominates at low coronal temperatures, and thermal conduction, that increases its efficiency at high coronal temperatures. 
\end{itemize}

Since the coronal temperature correlates with the virial mass of dark matter halo where galaxy resides, it appears that the ability of a galaxy to cool its own corona decreases for increasing the halo mass: it is high for $M_\mathrm{vir}\lesssim3\times10^{12}\,\mo$ and it is null for $M_\mathrm{vir}>10^{13}\,\mo$. We speculate that coronal condensation driven by interaction between fountain clouds and hot coronal gas could have important implications for galaxy evolution. It could have been a viable mechanism to sustain the star formation in Milky-Way-like galaxies after the turn-off of the \textit{cold mode} accretion. At the same time, its inefficiency in massive galaxies could provide a possible explanation for both mass and environment quenching of the star formation.

\section*{Acknowledgements}
We thank the anonymous referee for helpful suggestions and very constructive comments.
LA and FF thank Fabrizio Brighenti for useful suggestions concerning the implementation of thermal conduction in ATHENA. 
LA is grateful to Enrico Di Teodoro and Gabriele Pezzulli for their help. We acknowledge the CINECA award under the ISCRA initiative, for the availability of high performance computing resources and support. We acknowledge the Kapteyn Astronomical Institute to allow us to use the Gemini clusters to test the new algorithms included in ATHENA.

\bibliography{biblio}

\appendix
\section{Thermal conduction algorithm}\label{appendix}

\begin{figure*}
\centering
\includegraphics{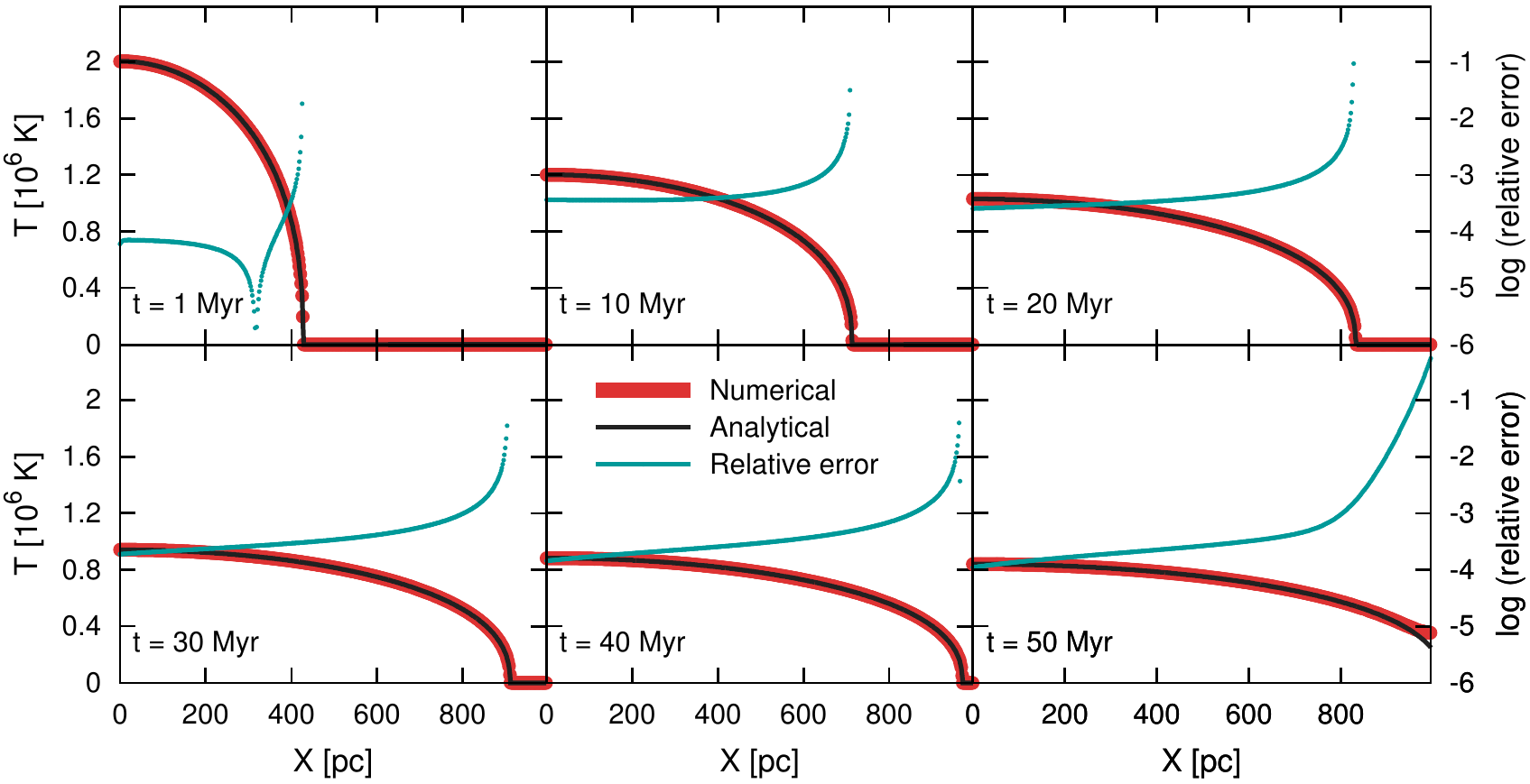}
\caption{Propagation of the conduction front with time. Comparison of the numerical results (red line) with the analitical solution (black line). Additionally the relative error between the two solution is plotted (cyan line).}
\label{test1}
\end{figure*} 

The effects of thermal conduction were treated by solving the heat conduction equation (Eq.~\ref{Spitzer}), a parabolic partial differential equation, that in a 2D cartesian geometry can be written:
\begin{equation}
\dfrac{\partial T}{\partial t} =\dfrac{T}{e} \left[ \frac{\partial}{\partial x} \left( \kappa_\mathrm{Sp} \frac{\partial T}{\partial x} \right) +  \frac{\partial}{\partial y}  \left( \kappa_\mathrm{Sp} \frac{\partial T}{\partial y} \right) \right] 
\label{heat}
\end{equation}
with $ e = P / (\gamma - 1) $ is the internal energy density of the gas and  $\gamma=5/3$. 

To solve the heat equation (Eq. \ref{heat}), we used an Alternating Direction Implicit (ADI) method, that splits the finite difference equations into the two spatial directions. Below we show the algorithm for the \textit{x} spatial direction, where the \textit{y} coordinate is kept constant. The \textit{y} spatial direction is treated separately using the same strategy, but with the new temperature distribution. We point out the non-linearity of the heat equation (Eq. \ref{heat}): $\kappa_\mathrm{Sp}$ is not constant but strongly temperature dependent,  $\kappa_\mathrm{Sp} \equiv \kappa_\mathrm{Sp}(T) \propto T^{5/2}$ (see Eq. \ref{Spitzer2}). We solved Eq. \ref{heat} using a specific method for non-linear parabolic partial differential equations \citep{NumericalRecipes}. Moreover, in order to not limit the hydro-timestep to the conduction-timestep and to avoid very long computational times, we used a semi-implicit time integration method. The approach was taken from \citet{Vieser&Hensler07}: the implicit part is weighted by a factor $0.5 < \alpha < 1$ and the explicit part by a factor $1-\alpha$. In pratice, we discretize Eq. \ref{heat} as follows:

\begin{equation}
\begin{split}
	& \dfrac{T_{i,j} ^{n+1} - T_{i,j} ^{n}}{\Delta t / 2}  \,  =  \alpha \, \, \,\left\lbrace \dfrac{T_{i,j} ^{n+1}}{e_{i,j} ^{n+1}}  \dfrac{1}{\Delta x^2} \,  \left( z_{i+1,j} ^{n+1} - 2z_{i,j} ^{n+1} +z_{i-1,j} ^{n+1} \right) \right\rbrace +(1 -\alpha) \,\, \\
	\vphantom{\Bigg|}	
	&  \, \left\lbrace \dfrac{T_{i,j} ^{n}}{e_{i,j} ^{n}}  \dfrac{1}{\Delta x^2} \, \left[ \kappa_\mathrm{Sp}(T_{i+1/2,j} ^{n}) \left( T_{i+1,j} ^{n} - T_{i,j} ^{n} \right) - \kappa_\mathrm{Sp}(T_{i-1/2,j} ^{n}) \left( T_{i,j} ^{n} - T_{i-1,j} ^{n} \right) \right]\right\rbrace
\end{split}	
\label{discr1}
\end{equation}
with 
\begin{equation}
dz=\kappa_\mathrm{Sp}(T)dT \,\Longrightarrow\,z=\int \kappa_\mathrm{Sp} (T) dT = \frac{2}{7} \kappa_\mathrm{Sp} (T) \, T \:,
\label{discr2}
\end{equation}
discretizing and expanding $z_{i,j} ^{n+1}$ as
\begin{equation}
z_{i,j} ^{n+1} \equiv z(T_{i,j} ^{n}) + ( T_{i,j} ^{n+1} - T_{i,j} ^{n}) \, \dfrac{\partial z}{\partial T}_{i,j} ^{n} = \frac{2}{7} \kappa_\mathrm{Sp}(T_{i,j} ^{n}) \, T_{i,j} ^{n} + ( T_{i,j} ^{n+1} - T_{i,j} ^{n}) \, \kappa_\mathrm{Sp}(T_{i,j} ^{n}) \:.
\label{discr3}
\end{equation}
\\
The equations \ref{discr1}-\ref{discr2}-\ref{discr3} can be rearranged in order to assume the following tridiagonal matrix form:
\begin{equation}
A_{i,j} T_{i-1,j} ^{n+1} + B_{i,j} T_{i,j} ^{n+1} + C_{i,j} T_{i+1,j} ^{n+1} = D_{i,j} \, ,
\label{matrix}
\end{equation}
where the \textit{{i}} index varies along the \textit{x} axis and the \textit{j} index is kept constant. This equation matrix can be efficiently solved with the `Thomas algorithm' \citep[see e.g.][]{NumericalRecipes}.
\\
Finally, the $\alpha$ parameter depends on the ratio between the hydro-timestep and the conduction-timestep, $\beta=\tau_{hydro}/\tau_{cond}$:
\begin{equation}
\alpha =\dfrac{1-\beta - \mathrm{exp}(-\beta)}{\beta (\mathrm{exp}(- \beta )-1)}
\end{equation}
with the additional limitation: $\alpha=0.5$ if $\alpha<0.5$ and $\alpha=1.0$ if $\alpha>1.0$. Therefore the implicit part contribution is dominant if the conduction-timestep is much smaller than the  hydro-timestep, and it is 50$\%$ or more in the other cases.

We have shown only the algorithm to solve Eq.~\ref{Spitzer}. The solution of Eq.~\ref{Eqcode} is similar but with $f [\kappa_\mathrm{Sp} / (1+\sigma)]$ instead of $\kappa_\mathrm{Sp}$. We treated $f [\kappa_\mathrm{Sp} / (1+\sigma)]$ as a single coefficient depending on temperature and we discretized $\sigma$ as follows:
\begin{equation}
\sigma = \dfrac{\kappa_\mathrm{Sp} || \nabla T||}{5 \Phi_{\mathrm{s}}\rho c^3} \,\Longrightarrow\, \sigma_{i,j} = \dfrac{\kappa_\mathrm{Sp}(T_{i,j}) \, ||(T_{i+1,j}-T_{i-1,j})/(2\Delta x)||}{5 \Phi_{\mathrm{s}}\rho_{i,j} {c_{i,j}}^3} \:,
\label{discr2}
\end{equation}
when solving the heat equation \ref{heat} in the x spatial direction. An analogous formula is used for the y direction. 

To test the accuracy of the algorithm we computed the numerical solution in the case of the classic propagation of a plane conduction front in a static uniform medium, in which the gas density is kept constant throughout the whole time evolution. For this problem an analytic solution for the temperature distribution as a function of time exists \citep{Reale95}:
\begin{subequations}
\begin{equation}
T = T_\mathrm{c}\,\left( 1-\dfrac{x^2}{x_\mathrm{f} ^2}\right) ^{2/5} \;,
\end{equation}
\begin{equation}
T_\mathrm{c} =0.6 \, \left(\dfrac{Q^2}{a t} \right) ^ {2/9} \;,
\label{A7b}
\end{equation}
\begin{equation}
 x_\mathrm{f} = 1.01 \, (Q^{5/2} a t)^{2/9} \;.
\label{A7c}
\end{equation}
\end{subequations}
For the numerical solution we took as initial profile the analytic solution at $t = 1$ Myr for a plasma with numeric density $n = 10^{-3}$ cm$^{-3}$. In \ref{A7b} \& \ref{A7c}, \textit{a} is a constant value equal to $5.6 \times 10^{-7}(\gamma - 1)/(k_\mathrm{B} n)$, with $ k_\mathrm{B} $ the Boltzmann constant. \textit{Q} is the integral of \textit{T} over the whole space, that we assumed to be $1.5 \times 10^{9}$ K pc, corresponding to an maximum temperature $T = 2 \times 10^6$ K at $t = 1$ Myr. The grid is 1 kpc long and it is composed by 500 cells, with a resolution of 2 pc, the same used in our fiducial simulations (see Sec. \ref{Numerical simulations}). Zero-gradient conditions on the temperature were imposed at all boundaries. Fig. \ref{test1} shows the numerical solution, the analytic solution and the relative error for different times. The numerical solution is in very good agreement with the analytic solution. The relative error within the hottest region of the plasma is less than $0.01 \%$. In the panels at 1, 10, 20, 30, 40 Myr the relative error trend becomes asymptotic when the temperature falls to zero. In this region the relative error is meaningless because it was calculated by dividing the difference between the analytic and numeric solution for a null value. In the last panel the difference between the two trends at the boundaries is visible and the relative error amounts to about $70 \%$. This is due to the temperature gradient forced to be zero at the grid boundary. With this test we could verify the accuracy of our implementation of thermal conduction in the code.

\label{lastpage}

\end{document}